# Simulation Study of Fast Ion Instability in the ILC Damping Ring and PETRA III*

G. Xia[1], K. Ohmi[2], E. Elsen[1]

[1]DESY, Hamurg, 22607, Germany
[2] KEK, 1-1 Oho, Tsukuba, 305-0801, Japan

## Abstract

The fast ion instability is simulated in different gas pressures and fill patterns for the damping ring of the International Linear Collider (ILC) and PETRA III respectively. Beam size variation due to beta function and dispersion function change is taken into account. Feedback is also applied in the simulation.

* This work is supported by the Commission of the European Communities under the 6th Framework Programme "Structuring the European Research Area", contract number RIDS-011899.

## 1 INTRODUCTION

Fast ion instability (FII) in electron storage rings is one of the two-stream instabilities. It is similar to electron cloud instability in positively charged particle storage rings. The ions can disturb the beam motion and lead to beam size blow up and tune shifts. This phenomenon was firstly observed in ALS ring and then confirmed by other rings such as PLS, TRISTAN-AR, ATF etc [1]. It is of importance for the performance of very low emittance storage rings like ILC damping ring and ring-based PETRA III light source.

This note is structured as follows. In section 2, the simulation procedure is briefly described. Simulation results of FII for the ILC damping ring are presented in section 3 by using the different fill patterns and in different gas pressures. The feedback system with damping time of 50 turns is also applied in the simulation. Section 4 gives the simulation results of FII for PETRA III in which the two operation modes are taken into account. In the end, a short summary is given.

## 2 SIMULATION STUDY

This is a weak-strong simulation code dedicated to FII study [2]. In this simulation, the electron bunch is treated as rigid Gaussian beam. Only the centroid motion of bunch is taken into account. The ions come from collisional ionization process and they are regarded as marco-particles. There is limited number of ionization points along the ring. The motion of ions is non-relativistic without longitudinal drift and they are assumed to drift freely in the bunch interval. The interaction between beam and ions is based on Bassetti-Erskine formula [3]. To connect the adjacent interaction points, the linear transport matrix are used. In this code, the beam size variation due to beta function and dispersion function change is taken into consideration.

The interaction between the beam and ions can be described in analogy with the beam-beam interaction. The velocity change of an ion is given by

$$\Delta v_{y,i} + i\Delta v_{x,i} = -2N_0 r_e c f(x_i, y_i) m_e / M_i \tag{1}$$

where $N_0$ is the number of electrons per bunch, $r_e$ is the classical radius of electron, $c$ is the speed of light, $m_e$ and $M_i$ are the rest mass of electron and ion, respectively.. $f(x,y)$ is the well known Bassetti-Erskine formula which is given by

$$f(x,y) = -\frac{\sqrt{\pi}}{\sqrt{2(\sigma_x^2 - \sigma_y^2)}}\left[ w\left(\frac{x+iy}{\sqrt{2(\sigma_x{}^2 - \sigma_y{}^2)}}\right) - \exp\left(-\frac{x^2}{2\sigma_x{}^2} - \frac{y^2}{2\sigma_y{}^2}\right) w\left(\frac{\frac{x\sigma_y}{\sigma_x} + i\frac{y\sigma_x}{\sigma_y}}{\sqrt{2(\sigma_x{}^2 - \sigma_y{}^2)}}\right)\right] \tag{2}$$

here

$$w(z) = \exp(-z^2)[1 - erf(-iz)] \tag{3}$$

and the error function is

$$erf(x) = \frac{2}{\sqrt{\pi}}\int_0^{\pi} \exp(-x^2)dx \tag{4}$$

where $x_i$, $y_i$ denote the transverse distances of ions with respect to the bunch centre, $\sigma_x, \sigma_y$ the transverse beam sizes. Therefore, the kick to the rigid electron by ion with distance $(x_{ie}, y_{ie})$ and summing for all the ions is expressed by

$$\Delta y'_e + i\Delta x'_e = \frac{2N_0 r_e}{\gamma} \sum_i f(x_{ie}, y_{ie}) \tag{5}$$

where $\gamma$ is the relativistic factor of the electron. Similarly, the kick to an ion with mass $M_i$ is given by

$$\Delta y'_i + i\Delta x'_i = -2N_0 r_e c \frac{m_e}{M_i} f(x_{ie}, y_{ie}) \tag{6}$$

where $(\Delta x'_e, \Delta y'_e)$ and $(\Delta x'_i, \Delta y'_i)$ are the transverse angle kicks to the centre-of-mass of electron bunch and ions respectively.

The number of ionization points is chosen to be equal to the number of optical elements of the accelerator lattice. In order to save computation time, our simulation uses one of the symmetric sections of the ring as the interaction region between the beam and ions. In other sections of the ring there is no ion production. New marco-particles are produced at the location of beam position *(x, x′, y, y′)* where the ionization occurs. The motion of the beam and ion is tracked from turn to turn. Since the vertical beam emittance is smaller than the horizontal one (the emittance coupling ratio is about 0.4 % and 1.0 % for ILC damping ring and PETRA III respectively), the FII is much serious in the vertical plane.

The dipole moment of each bunch is computed and recorded in every turn. The vertical amplitude of bunch centroid is half of the Courant-Synder invariant which is given by

$$J_y = \frac{1}{2}\left[\frac{(1+\alpha^2)}{\beta} y^2 + 2\alpha y y' + \beta y'^2\right] \tag{7}$$

where $\alpha$ and $\beta$ are the Twiss parameters which depend on the optical design of the ring. We compare $\sqrt{J_y}$ with the beam size which is represented by $\sqrt{\varepsilon_y}$ , here $\varepsilon_y$ is the vertical emittance of the beam. Both of these quantities are in units of $m^{1/2}$.

## 3 SIMULATION RESULTS FOR ILC DAMPING RING

One of sextant of the ILC damping ring lattice is chosen in the simulation. The basic beam parameters are listed in Table 1 for two fill pattern case *A* and *B*. We assume the ionization cross section 2 Mbarn for CO with a molecular mass number of 28. The partial gas pressure is changed from 0.2 nTorr to 1.0 nTorr. Since the ion density for mini-trains can quickly reach the peak value after the first few bunch trains. 5 & 10 bunch trains are assumed in the simulation [4]. Growth time of FII in different fill pattern cases and different gas pressures can be estimated from the simulation results (The growth time of FII indicates the time duration of maximum amplitude growth of beam from 0.1 $\sigma_y$ to 1.0 $\sigma_y$).

Table 1: Two typical fill patterns in the ILC damping ring.

| *Fill patterns* | *A* | *B* |
|---|---|---|
| Number of bunches, $n_b$ | 4346 | 2767 |
| Particles per bunch, $N_0$ [$10^{10}$] | 1.29 | 2.02 |
| Bunch spacing, [bucket] | 2 | 4 |
| Number of trains, $p$ | 82 | 61 |
| Bunches per train, $f_2$ [bucket] | 0 | 23 |
| Gap between trains, $g_2$ [bucket] | 0 | 28 |
| Bunches per train, $f_1$ [bucket] | 53 | 22 |
| Gap between trains, $g_1$ [bucket] | 71 | 28 |

Figure 1 and Figure 2 give the evolution of the maximum vertical amplitude of bunch centroid with respect to number of turns for fill pattern *A* in CO pressure of 0.2nTorr without and with feedback system, respectively. Some notations used here are listed as following. $N_0$ is the number of particles per bunch, $n_b$ is the number of bunches per train, $n_{train}$ is the train number, Lsep is the bunch spacing in units of RF bucket, LtrainGap is the gap length between two adjacent bunch trains. The estimated growth time is also shown here. It can be seen that feedback system with the damping time of 50 turns can damp the FII in this case. Similarly, Figure 3 and Figure 4 give the evolution of the maximum vertical amplitude in CO pressure of 1nTorr. It indicates that even by using fast feedback system, the FII can not be totally damped. Figure 5 and 6 give the simulation results in 1nTorr for fill pattern *B*. It shows FII grows faster for longer bunch trains (46 bunches per train) comparing to shorter one (23 bunches per train). This phenomenon can also be clearly shown in Figure 7 in which we use 5 trains with 46 bunches per train and 10 trains with 23 bunches per train respectively (the total number of bunches is the same in both cases). Figure 8 and Figure 9 show the beam centroid oscillation pattern in different turns without and with feedback system for fill pattern *B* in CO pressure of 1 nTorr. It indicates that the beam oscillation grows with respect to the time (number of turns). Meanwhile, the tail bunch oscillates with larger amplitude than that of the head bunch. This is also one of the characteristics of FII. It can be seen that by introducing the fast feedback system, the beam oscillation can be greatly reduced.

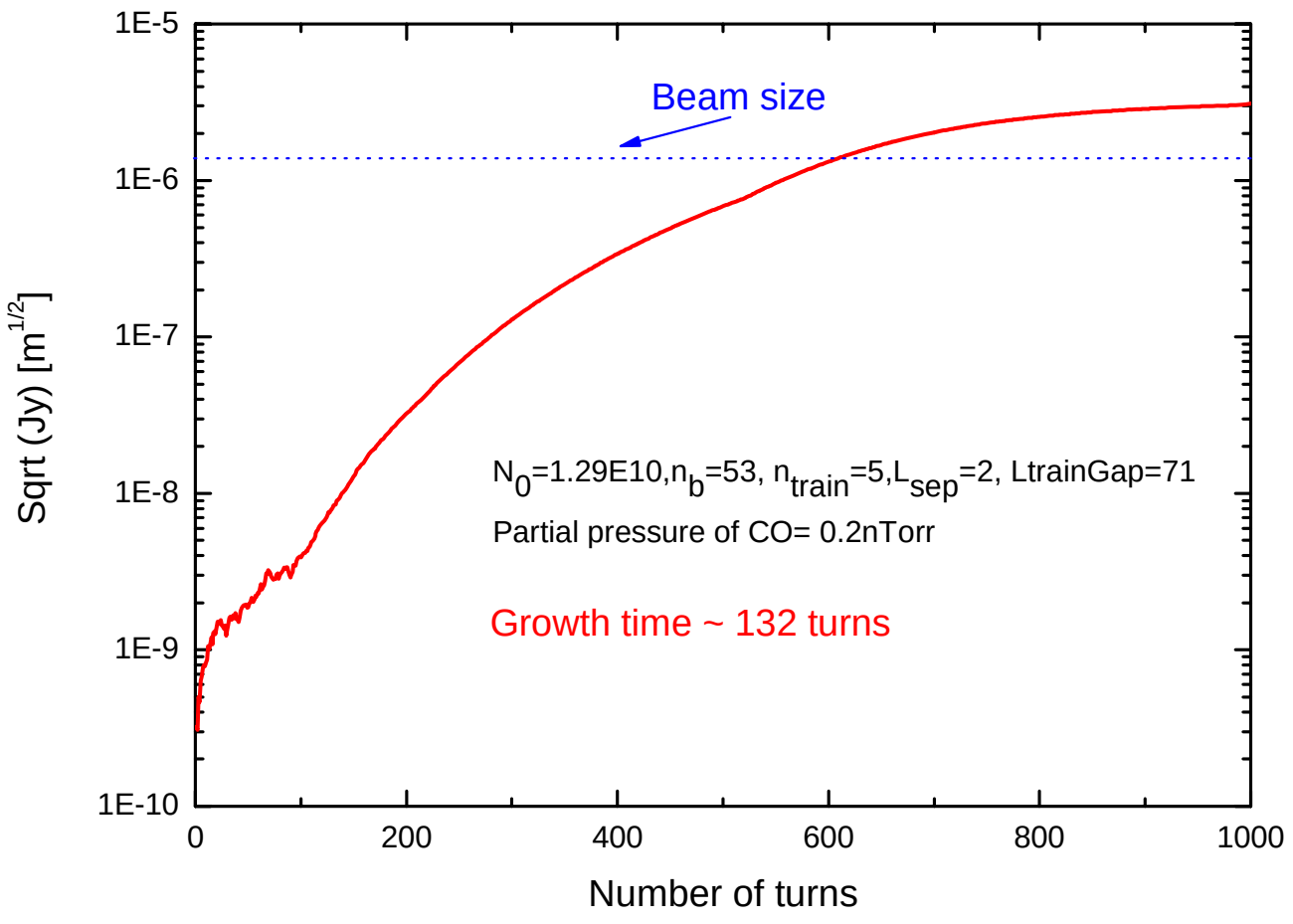

Fig.1: Evolution of the maximum amplitude for fill *A* in 0.2nTorr without feedback.

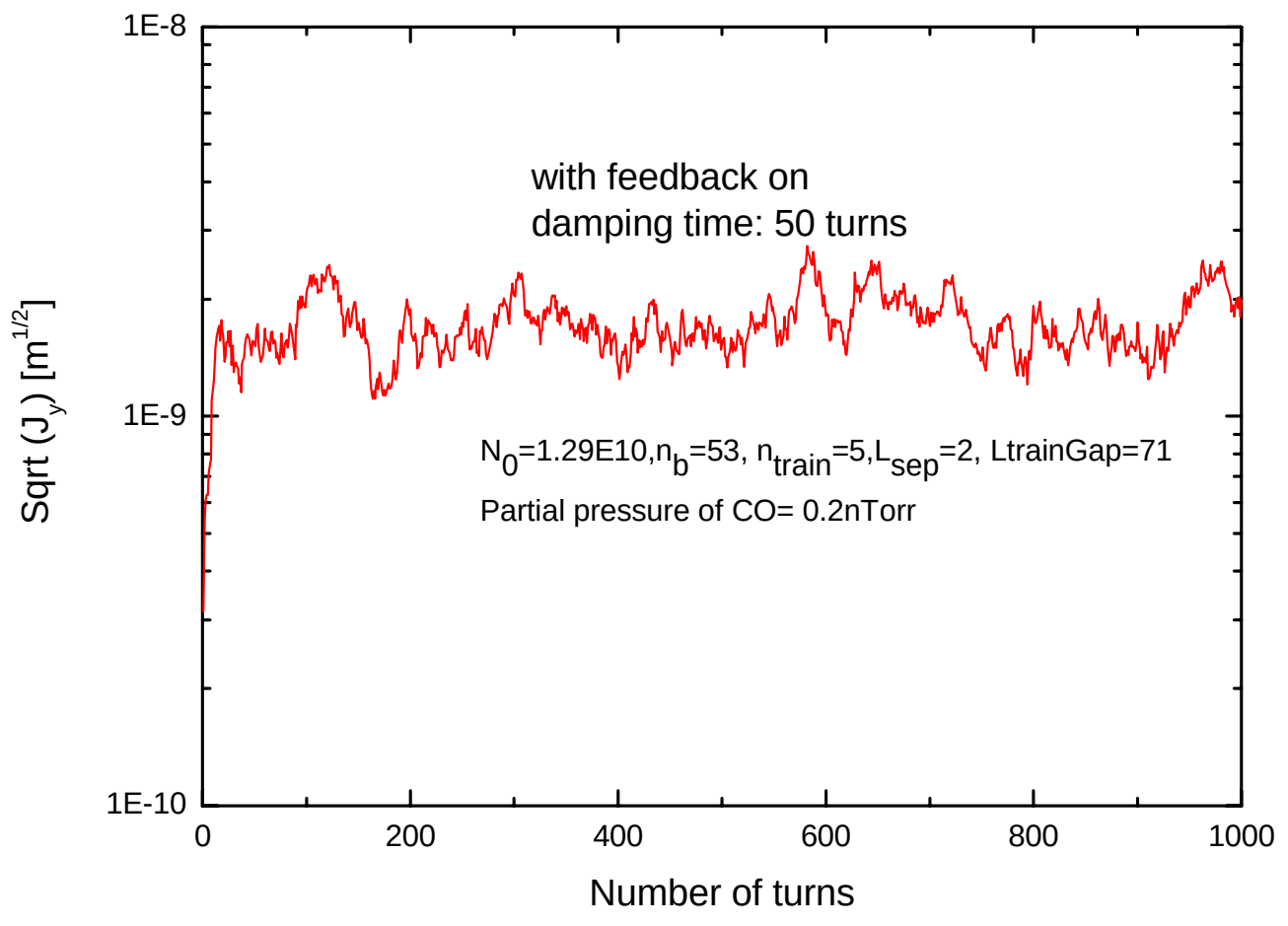

Fig.2: Evolution of the maximum amplitude for fill *A* in 0.2nTorr with feedback on.

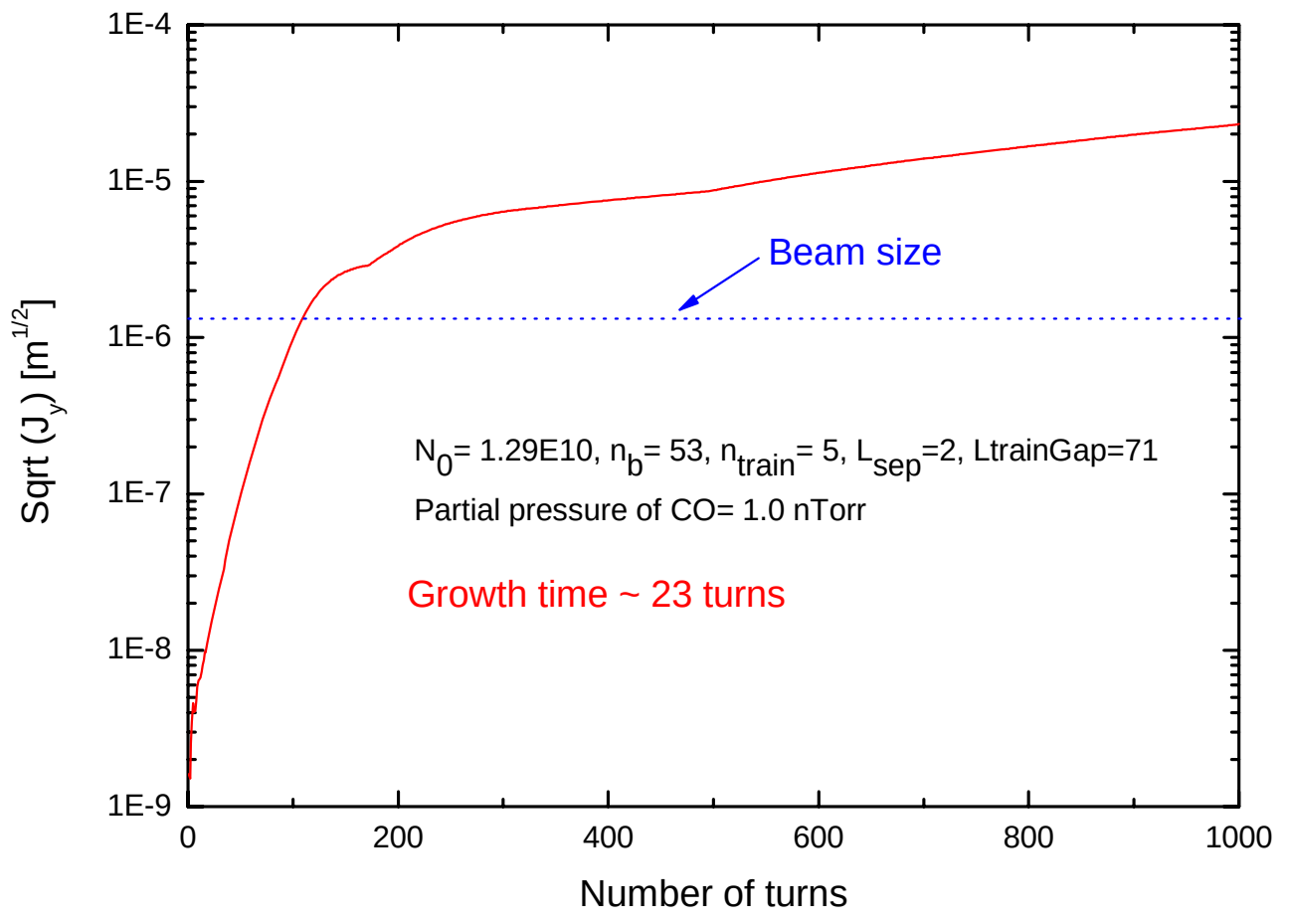

Fig.3: Evolution of the maximum amplitude for fill *A* in 1.0nTorr without feedback on.

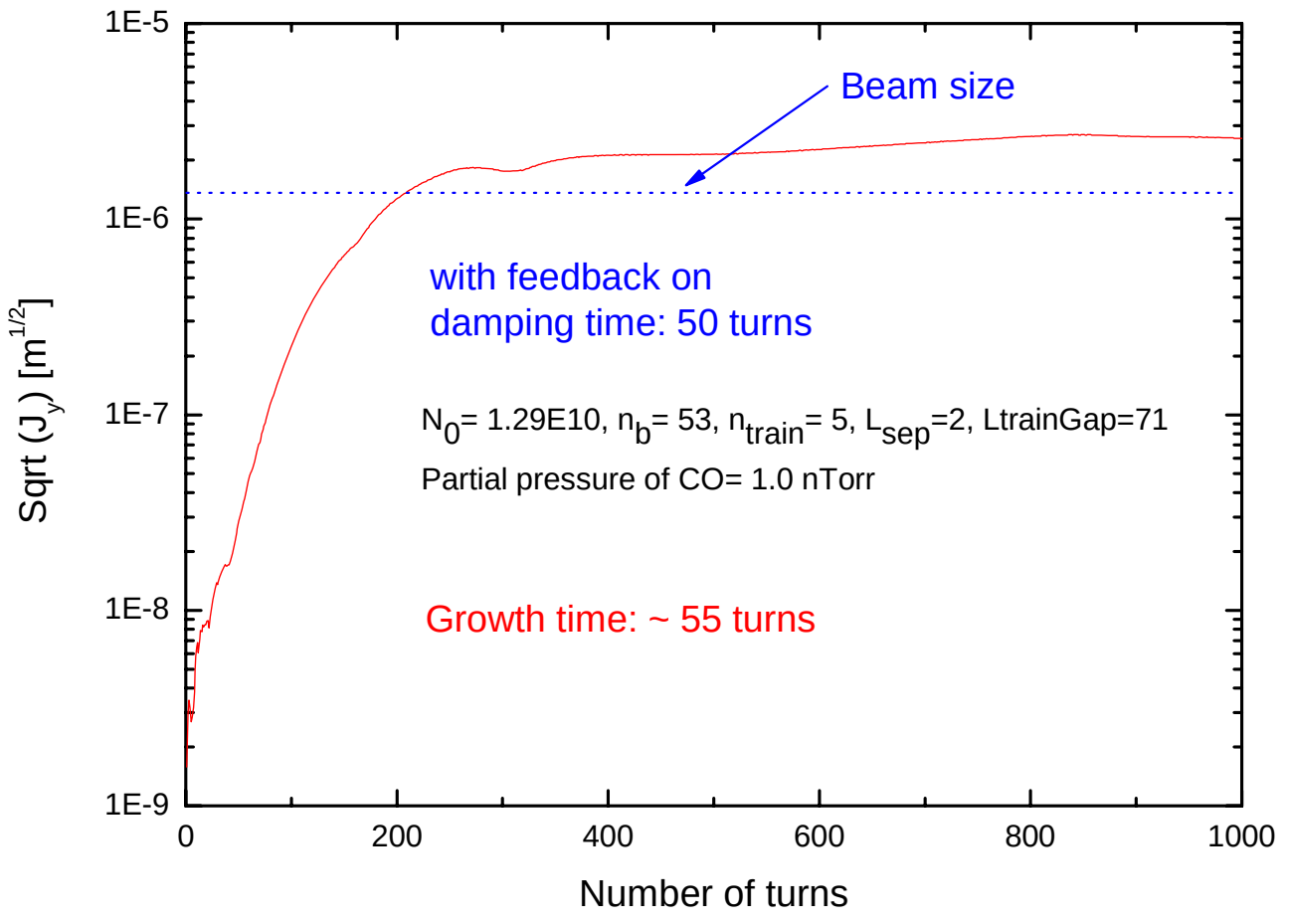

Fig.4: Evolution of the maximum amplitude for fill *A* in 1.0nTorr with feedback on.

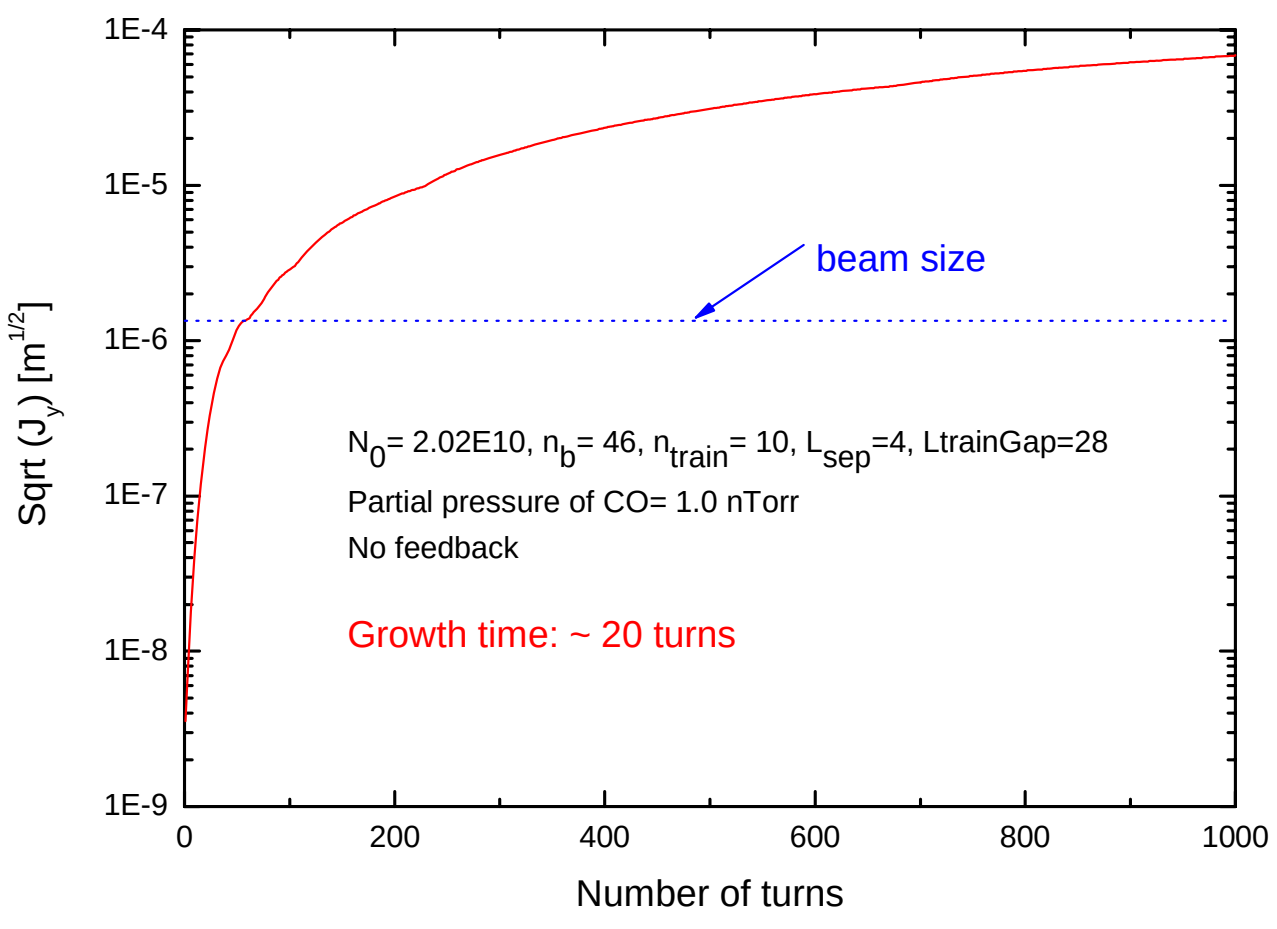

Fig.5: Evolution of the maximum amplitude for fill *B* in 1nTorr without feedback on.

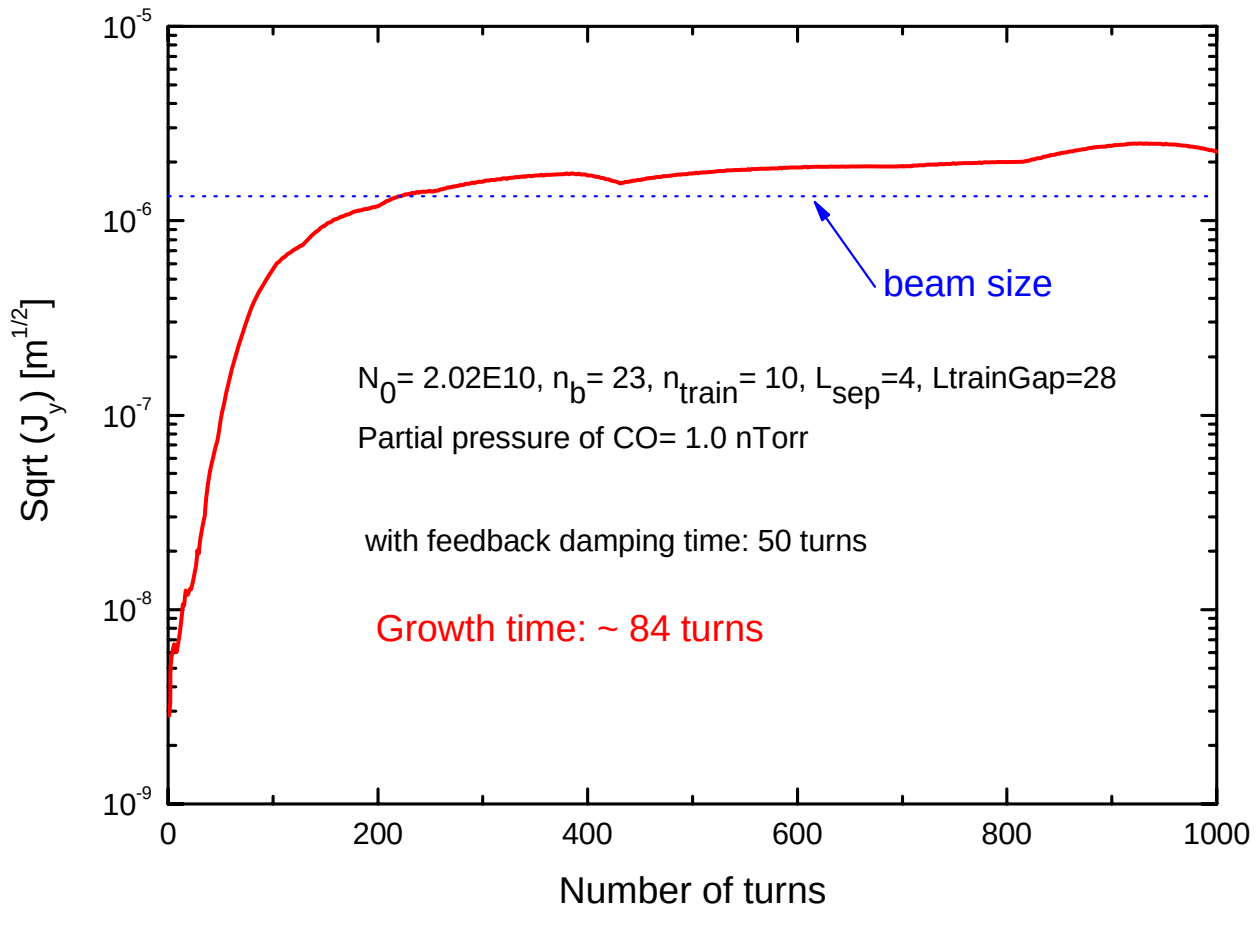

Fig.6: Evolution of the maximum amplitude for fill *B* in 1nTorr with feedback on.

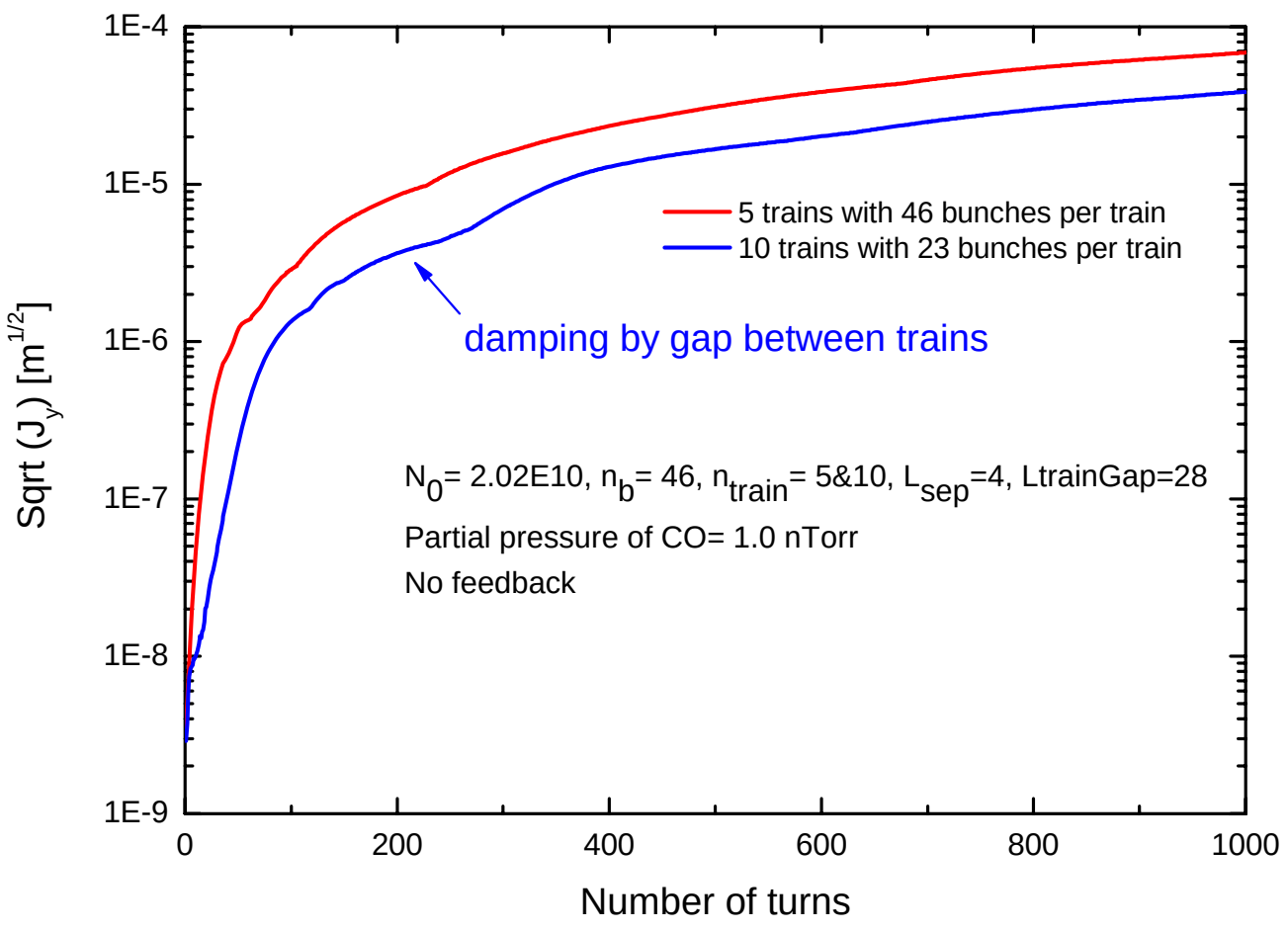

Fig.7: Evolution of the maximum amplitude for fill *B* in 1nTorr for two kinds of mini-trains.

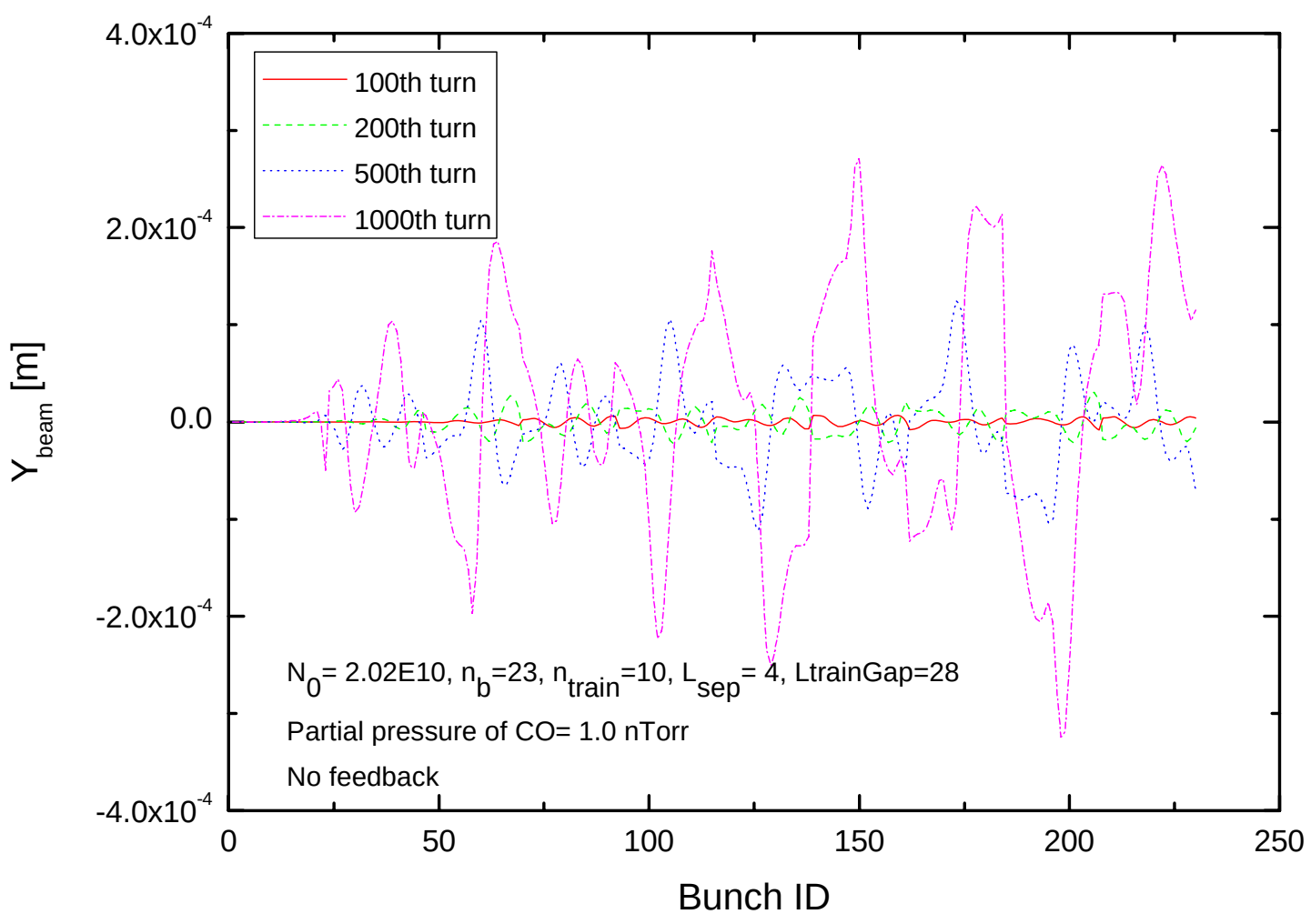

Fig.8: Beam oscillation pattern in different turns for fill *B* without feedback on.

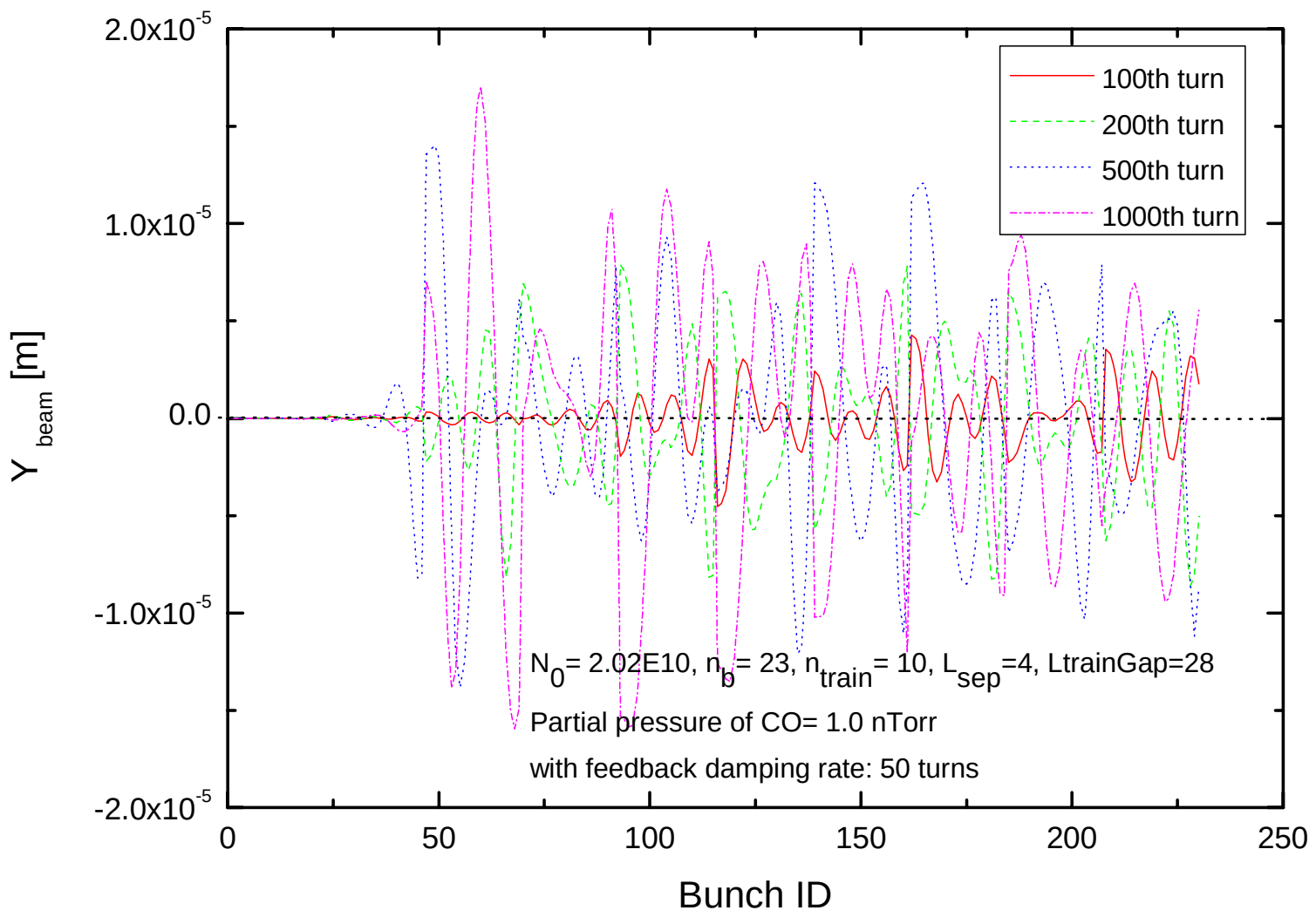

Fig.9: Beam oscillation pattern in different turns for fill *B* with feedback on.

## 4 SIMULATION RESULTS FOR PETRA III

PETRA III is a low emittance storage ring dedicated to synchrotron radiation on DESY site in Hamburg. The main parameters of PETRA III are listed in Table 2. There are two operation modes in this ring. One is multi bunch mode and the other is time resolved mode [5]. These bunches in both cases are evenly distributed around the ring. We simulate the FII in these two modes respectively and the results are shown in Figure 10 to Figure 16.

Table 2: Beam parameters of PETRAIII.

| Parameters | Value |
|---|---|
| Circumference [m] | 2304 |
| Harmonic number | 3840 |
| Energy [GeV] | 6 |
| RF [MHz] | 499.564 |
| Beam current [mA] | 100 |
| Total number of positrons [$10^{12}$] | 4.8 |
| Bunch number (multi bunch mode) | 960 |
| Bunch number (time resolved mode) | 40 |
| Horizontal emittance [nm] | 1 |
| Vertical emittance [nm] | 0.01 |
| Horizontal tune | 37.26 |
| Vertical tune | 33.20 |
| Synchrotron tune | 0.049 |
| Momentum compaction factor [$10^{-4}$] | 1.2 |
| Bunch length [mm] | 13.2 |
| Energy spread [$10^{-3}$] | 1.27 |

The evolution of the maximum vertical amplitude for multi bunch mode without feedback in CO gas pressure of 0.225 nTorr, 0.5 nTorr and 1.0 nTorr are given in Figure 10, 11 and 12 respectively. It can be seen that the oscillation amplitudes of beam will reach beyond the beam size in these cases. The estimated growth time of FII is also noted in each Figure. Taking into account the fast feedback system, the bunch oscillation amplitude is shown in Figure 13. It can be seen that the current feedback system with damping time of 50 turns can effectively damp the FII. Growth time of FII versus vacuum pressure is shown in Figure 14. We can see that the FII growth time is less for high vacuum pressure which is also shown in linear theory of FII [6]. Figure 15 and Figure 16 show the evolution of the maximum amplitude of beam for time resolved mode in CO pressure of 0.225nTorr and 1nTorr respectively. It can be seen that the bunch oscillation amplitude is well below the beam size. This is because in this case, the bunch spacing is very large which can drive the ions to large amplitude. These scattered ions form the ion halo and do not affect the beam motion seriously.

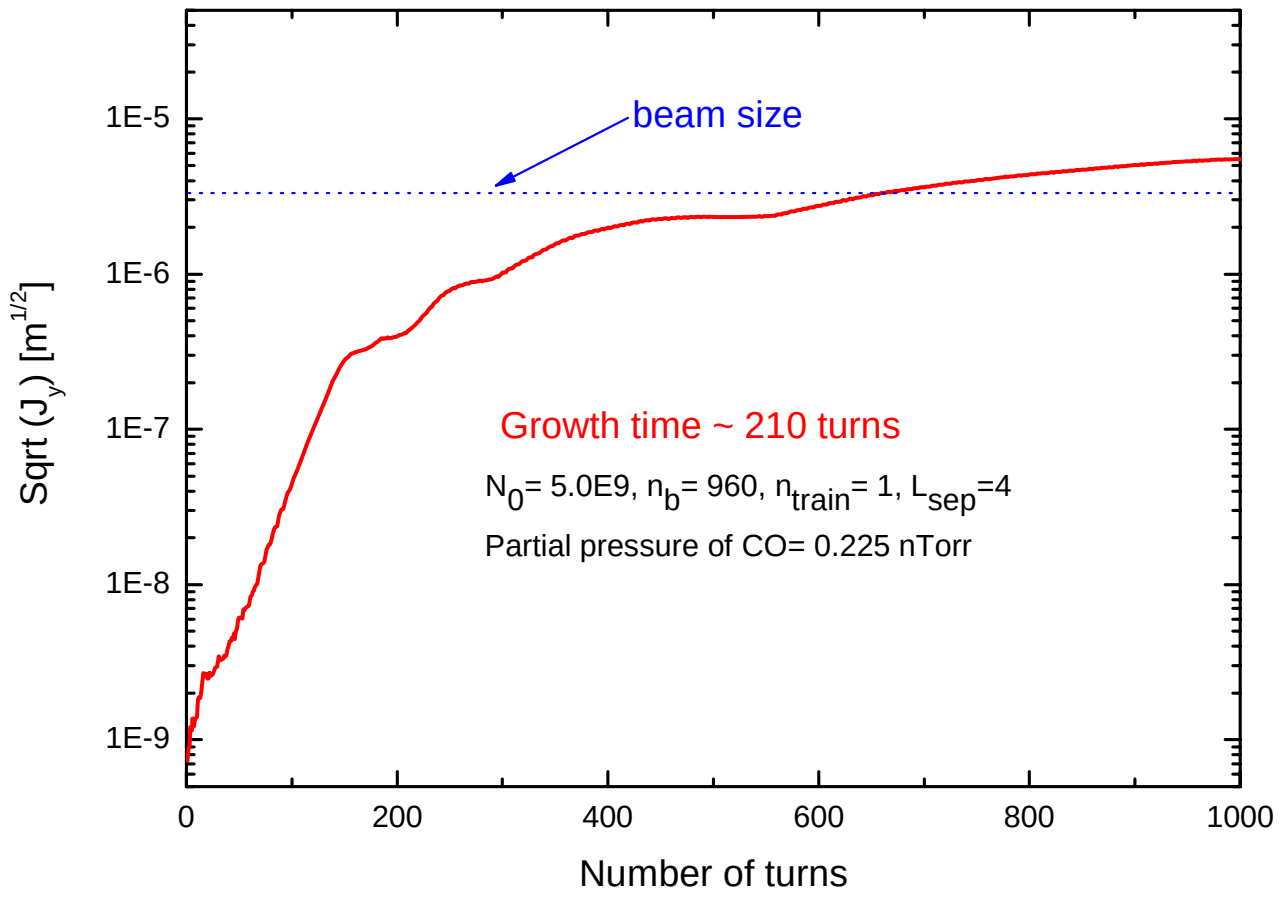

Fig.10: Evolution of the maximum amplitude for multi bunch mode in 0.225 nTorr without feedback on.

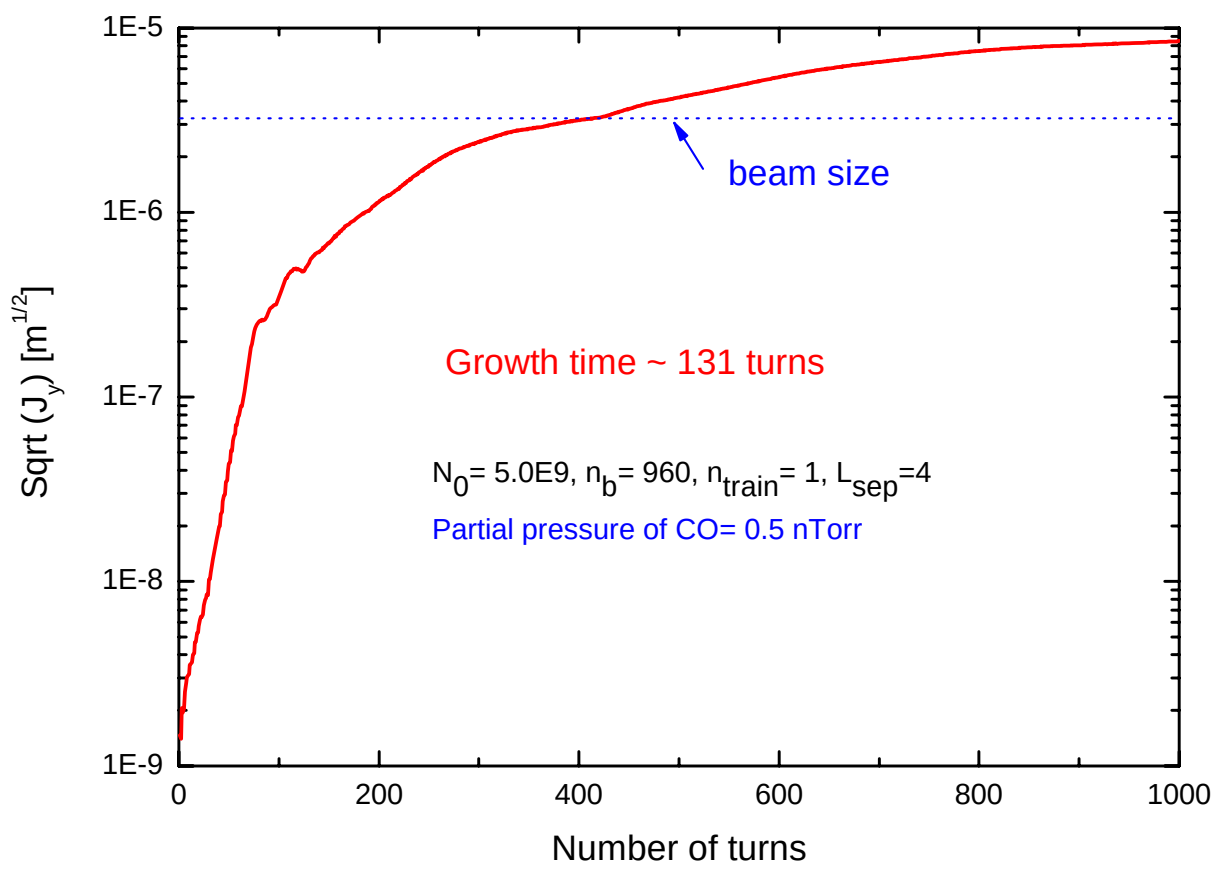

Fig.11: Evolution of the maximum amplitude for multi bunch mode in 0.5nTorr without feedback.

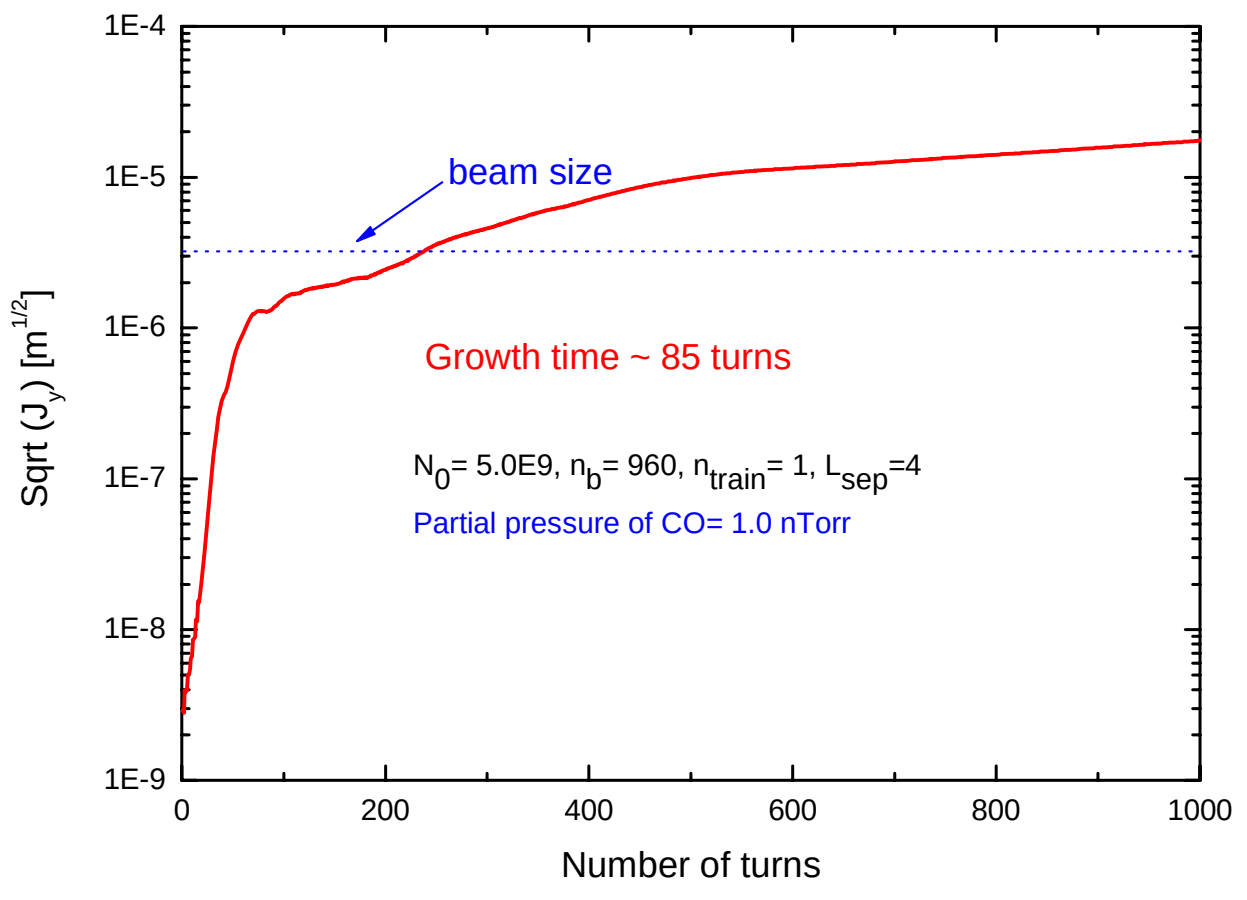

Fig.12: Evolution of the maximum amplitude for multi bunch mode in 1nTorr without feedback on.

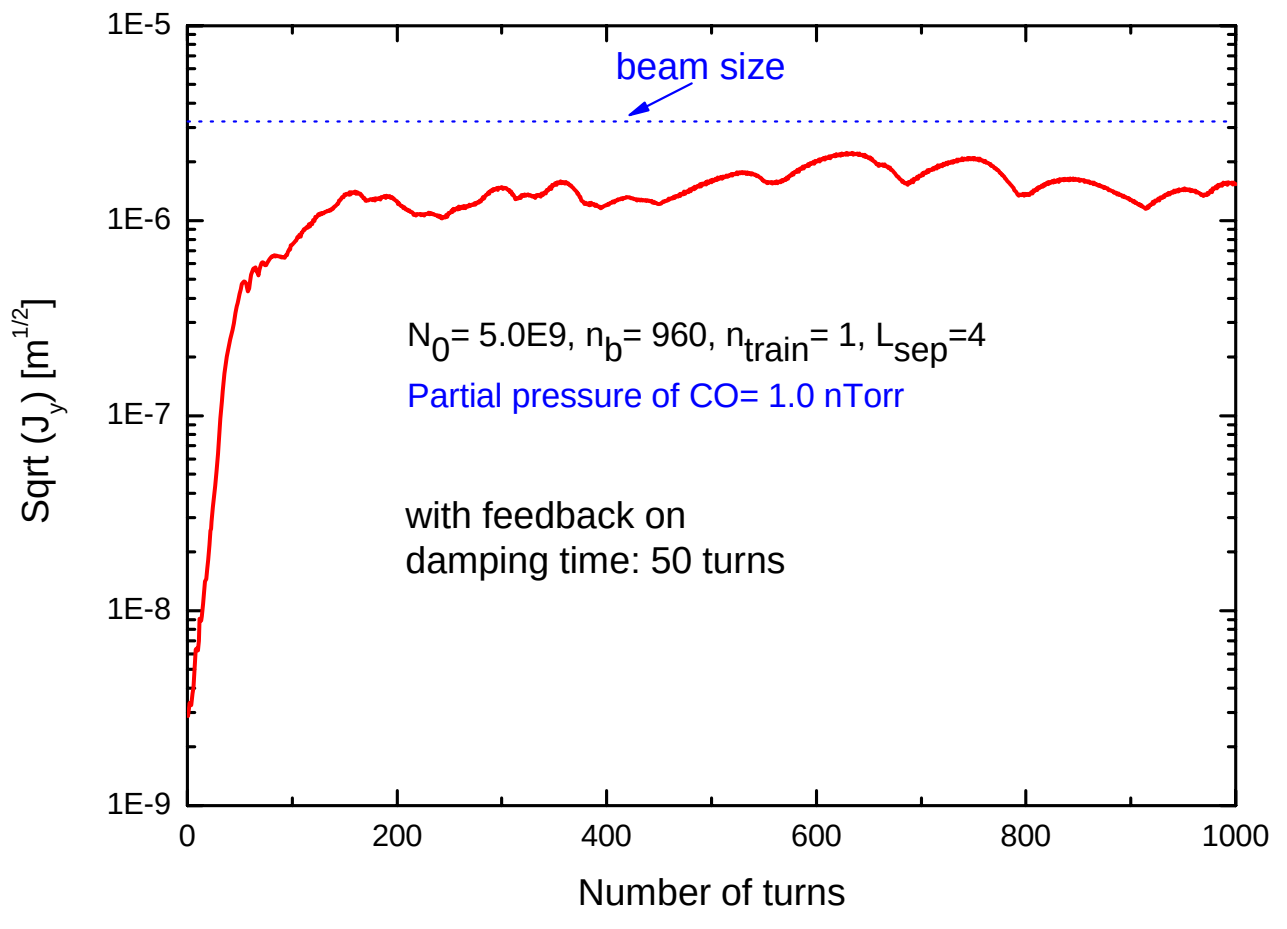

Fig.13: Evolution of the maximum amplitude for multi bunch mode in 1nTorr with feedback on.

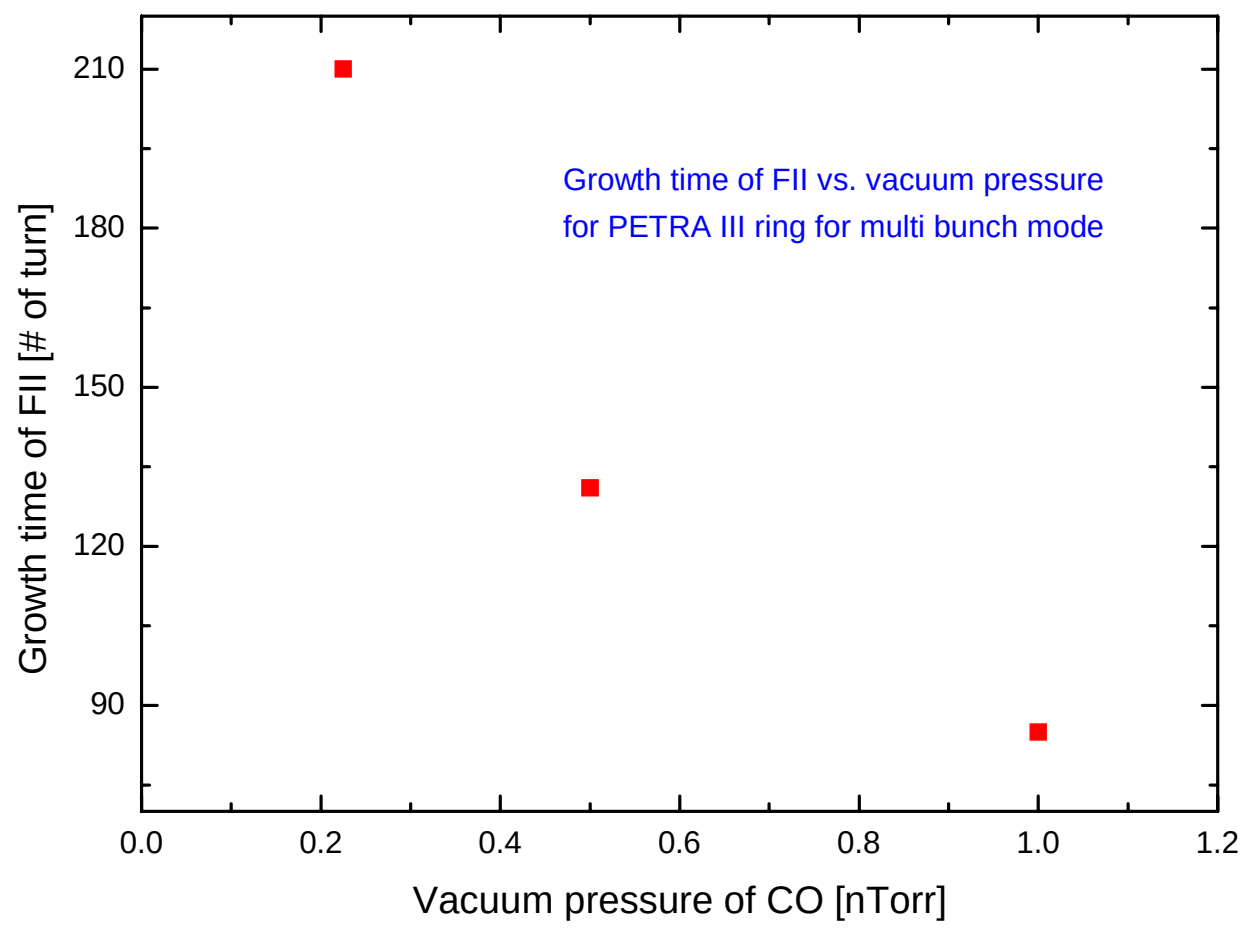

Fig.14: Growth time of FII *vs.* vacuum pressure.

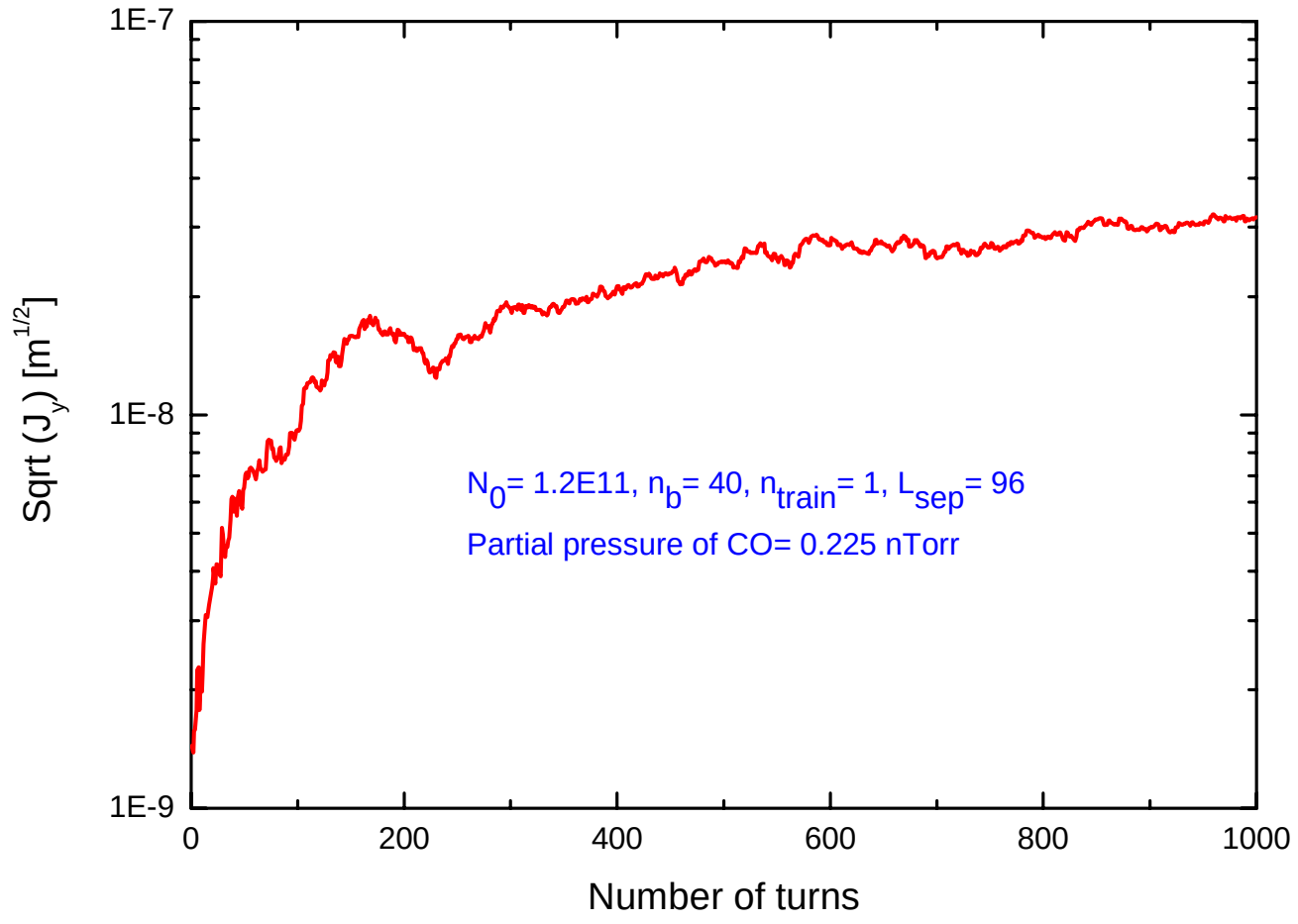

Fig.15: Evolution of the maximum amplitude for time resolved mode in 0.225nTorr without feedback.

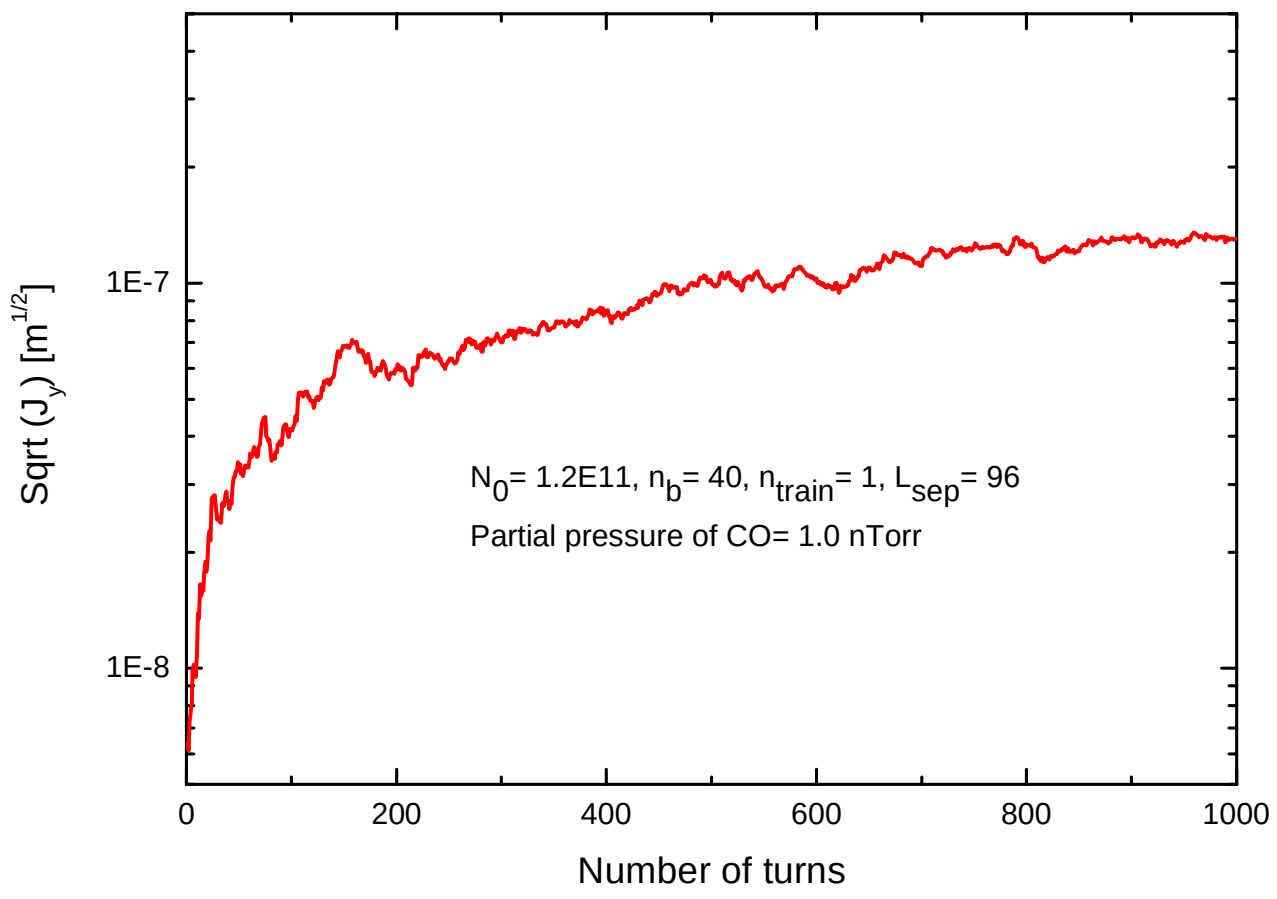

Fig.16: Evolution of the maximum amplitude for time resolved mode in 1.0 nTorr without feedback.

## 5 SUMMARY

The simulation results in this note indicate that in the typical fill pattern *A* and *B* for the ILC damping ring, the fast ion instability can not be damped by the bunch by bunch feedback system with the damping time of 50 turns if the gas pressure of CO is larger than 1 nTorr. Therefore, the lower gas pressure, namely, less than 1nTorr and even faster feedback system are critical to mitigate the FII. For the PETRA III ring, the FII will not affect beam seriously in time resolved mode, while for multi-bunch mode, the FII can lead to beam oscillation to large amplitude which is beyond the beam size. Therefore, the feedback system is necessary to damp the beam oscillation growth due to ions.